\documentclass[12pt,a4paper,oneside]{article}
\usepackage{graphicx}
\usepackage{color}
\usepackage{booktabs}
\usepackage{epsfig}
\usepackage{multirow}
\usepackage[english]{babel}
\usepackage{amsmath}
\usepackage{amssymb}
\usepackage[latin1]{inputenc}
\usepackage{fancyhdr}
\usepackage{indentfirst}
\usepackage{newlfont}
\usepackage{latexsym}
\usepackage[latin1]{inputenc}
\usepackage{fancyhdr}
\usepackage{booktabs}
\usepackage[hang,bf,small]{caption}
\usepackage{microtype}
\usepackage[a4paper,top=4cm,left=4cm]{geometry}
\title{WDVV Equations} 
\author{F. Magri\\
\normalsize{Dipartimento di Matematica e Applicazioni}\\
\normalsize{Università di Milano Bicocca}\\
\normalsize{franco.magri$@$unimib.it }\\
}
\date{}
\begin{document}
\maketitle 
\begin{abstract}
The paper aims to point out a novel geometric characterisation of the WDVV equations of 2D topological field theory.
\end{abstract}
\section{Introduction}
The Witten-Dijkgraaf-Verlinde-Verlinde equations are a remarkable set of nonlinear partial differential equations discovered in 2D topological field theory at the end of 80's \cite{1}. Afterwards they have found numerous and interesting applications in many areas. I simply quote the 2D SUSY Yang-Mills theory, the theory of Seiberg-Witten systems, the theory of Whitham equations, and the theory of integrable systems \cite{2}.

The mathematical structure of the WDVV equations has already been thoroughly studied, in the 90's, by Boris Dubrovin, who has invented the beautiful and far reaching concept of Frobenius manifold to give the WDVV equations a geometric interpretation. Since then, the theory of Frobenius manifolds has become a subject of interest in itself \cite{3}\cite{4}\cite{5}.
The purpose of the present paper is to consider again the question of the mathematical structure of the WDVV equations. I wish to argue that there are two distinct ways of dealing with these equations. On one hand, they can be seen as the equations defining a special class of associative and commutative algebras. This is the point of view followed by Boris Dubrovin, leading to the theory of Frobenius manifolds. On the other hand, the WDVV equations may be red as the equations defining special arrangements of 1-forms on a manifold, called Lenard complexes. This is the point of view worked out in this paper, leading to the theory of Haantjes manifolds. 

The paper is rather concise and direct, and the references to the theory of WDVV equations and to the theory of Frobenius manifolds are reduced to the bare essential. The WDVV equations are defined in Sec. 2. In Sec. 3 I remind the concept of Frobenius manifold and its relationship to the WDVV equations. In Sec. 4  I present the concept of ``Lenard complex'', and I explain why it is related to  the WDVV equations. 

\setcounter{equation}{0}
\renewcommand{\theequation}{\thesection.\arabic{equation}}

\section{WDVV equations}
The WDVV equations are an overdetermined system of nonlinear PDE's on a single function $F(x_1,x_2,...,x_n)$ of n coordinates. The equations are constructed in two steps. First one considers the Hessian matrix of the function F
\begin{equation}
h=Hessian(F) ,
\end{equation}
 and then the components
\begin{equation}
c_j=\frac{\partial h}{\partial x_j} .
\end{equation}
of the gradient of this matrix with respect to the given coordinates $(x_1,x_2,...,x_n)$. One of these matrices, say $c_1$, is assumed to be invertible. The WDVV equations are hence written in the matrix form
\begin{equation}
\label{5}
\frac{\partial h}{\partial x_j}\cdot \left( \frac{\partial h}{\partial x_1}\right)^{^{-1}}\cdot \frac{\partial h}{\partial x_l}=\frac{\partial h}{\partial x_l}\cdot \left( \frac{\partial h}{\partial x_1}\right)^{^{-1}}\cdot \frac{\partial h}{\partial x_j}.
\end{equation} 

Often a second assumption is made on the matrix $c_1$, by requiring that it does not depend on the coordinates $x_j$:
\begin{equation}
\label{4}
\frac{\partial c_1}{\partial x_j}=0.
\end{equation}
The acceptance of this assumption, however, depends on the field of interest. For instance, it is accepted in 2D topological field theory and in the theory of Whitham equations, but it is rejected in the theory of Seiberg-Witten systems \cite{6}. 
Accordingly I will consider Eq.(\ref{4}) as an auxiliary assumption, useful but not indispensable to define the WDVV equations. In other words, in this paper I consider the ``generalized'' WDVV equations, which abstract from condition (\ref{4}), contrary to the ``ordinary'' WDVV equations which require it \cite{7}\cite{8}.

There is another form of the WDVV equations which  must be quoted. It is
\begin{equation}
\label{6}
C_j\cdot C_l-C_l\cdot C_j=0 ,
\end{equation}
and concerns the matrices
\begin{equation}
\label{7}
C_j=\left( \frac{\partial h}{\partial x_1}\right)^{-1}\cdot \frac{\partial h}{\partial x_j}.
\end{equation}
In this form the WDVV equations are more specifically called ``associativity equations'', since they entail that the enties $C_{jk}^l(x_1,...,x_n)$ of the matrices $C_j$ are the ``structure constants'' of an associative commutative algebra with unity.

 The WDVV equations are clearly non tensorial. Any change of coordinates, different from an affine transformation, destroys the form of the equations. The coordinates  $(x_1,x_2,...,x_n)$ are therefore a basic constituent of the theory. This remark serves to clarify that the problem of giving the WDVV equations a geometric ( or intrinsic ) interpretation has two complementary aspects. On one side, one would write these equations in a form that does not depend on the choice of the coordinates. On the other side, one must demand that the geometrical structure used to attain this result should be capable to select a class of coordinates, affinely related, in which the intrinsic equations take the specific form $(2.3)$. The rule for the selection of the distinguished coordinates  $(x_1,x_2,...,x_n)$ is the key point of the process of geometrization of the WDVV equations. It is also the point which will mark the difference between the two approaches discussed in this paper.

  Before leaving the subject of WDVV equations, let me show an amazing example. It has been worked out by M. Kontsevich in 1994 \cite{9}. The equation 
\[f_{x_2x_2x_3}^2=f_{x_3x_3x_3}+f_{x_2x_2x_2}\cdot f_{x_2x_3x_3}\]
is one of the most simple instances of WDVV equations in $\mathbb{R}^3$. It comes from the standard WDVV equation $(2.3)$ by choosing  a function F of the form
\[F(x_1x_2x_3)=\frac{1}{2}(x_1^2x_3+x_1x_2^2)+f(x_2,x_3).\]
Albeit the equation is nonlinear, one may look for a solution in form of series
\[ F=\sum_{k}\frac{N_k}{(3k-1)\!}x_3^{3k-1}e^{kx_2}. \]
The insertion of this series into the equation gives a recursive relation on the coefficients:
\[ N_1=1 \qquad N_k=\sum_{p+q=k\ge 2}N_pN_qp^2q\left[q \binom{3k-4}{3k-2}-p\binom{3k-4}{3k-1}\right] \]
The first coefficients are
\begin{align}
N_2&=1 \nonumber\\
N_3&=12 \nonumber\\
N_4&=620 \nonumber
\end{align}
They are integers. These integers are of interest for the enumerative geometry.  Indeed, $N_1$ is the number of straightlines passing through two distinct points; $N_2$ is the number of conics passing through five points in generic positions; $N_3$ is the number of cubics passing through eight points, and so on \cite{10}. As proved by Kontsevich, all these coefficients are of interest to the enumerative geometry.

\setcounter{equation}{0}
\renewcommand{\theequation}{\thesection.\arabic{equation}}

\section{Frobenius manifolds}
In the early 90's, Boris Dubrovin has been the first to tackle the problem of giving the WDVV equations an intrinsic form, namely a form which does not depend on the choice of the coordinates. His idea has been to focus  the attention on  the matrices $c_1$ and $C_j$ which enter into the WDVV equations in the way explained before. He considers the  symmetric matrix $c_1$ as the matrix of the components of a semiriemmannian metric, in the distinguished coordinate system  $(x_1,...,x_n)$. Since he uses the additional assumption
\begin{equation}
\label{31}
\frac{\partial c_1}{\partial x_j}=0,
\end{equation}
characteristic of 2D topological field theory , this metric is flat. Furthermore, he considers the entries $C_{jk}^l(x_1,...,x_n)$ of the matrices $(C_1=Id,C_2,...,C_n)$ as the components of a third-order tensor field of type (1,2). This tensor field defines  a product on the tangent bundle. The  product is first defined on the basis associated with the coordinates $(x_1,...,x_n)$
\begin{equation}
\label{32}
\partial_j \circ\partial_k=C_{jk}^l\partial_l,
\end{equation}
and then  extended by linearity. It is known under the name of ``multiplicative structure on the tangent bundle''. If the matrices $C_j$ verify the associativity equations $(2.5)$ , this product is clearly  associative, commutative, and with unity. The unity is the distinguished vector field $\frac{\partial}{\partial x_1}$.

These preliminary remarks have led  Boris Dubrovin to choose, as the proper setting where to frame the geometric study of the WDVV equations, a class of manifolds endowed with three geometric structures:
\begin{align}
1. &\text{ a flat semi-riemannian metrics} \quad&g&:TM\times TM\rightarrow \mathbb{R} \nonumber\\
2. &\text{ an associative and commutative product} \quad&\circ&:TM\times TM\rightarrow TM \nonumber\\
\quad&\text{on the tangent bundle} \nonumber\\
3. &\text{ a distinguished vector field} \quad&e&:M\rightarrow TM. \nonumber
\end{align}
The conditions which must relate these structures in order to reproduce the WDVV equations are specified by Dubrovin  in his definition of Frobenius manifold.

\bigskip
\noindent
\emph{\textbf{Definition 1.}} \textit{A Frobenius manifold is a semiriemannian manifold (M,g), endowed with an associative and commutative product on the tangent bundle, and with a distinguished vector field $e:M\rightarrow TM$, which verify the following five conditions:
\begin{enumerate}
\item $Riemann(g)=0$
\item $g(X\circ Y,Z)=g(X,Y\circ Z)$
\item $e\circ X=X$
\item $\nabla_{\bullet}e=0$
\item $\nabla_{W}(g(X\circ Y,Z))=\nabla_{Z}(g(X\circ Y,W))$
\end{enumerate}
for any set of four covariantly constant vector fields (X,Y,Z,W).}

\bigskip
\noindent
The first four conditions are simple and natural: the metric is flat; the product is symmetric with respect to the metric; the distinguished vector field $e$ is the unity of the product; it is covariantly constant with respect to the Levi Civita connection induced by the metric. The fifth condition, on the contrary, is much less intuitive. It demands that the covariant derivative of the symmetric tensor field $g(X\circ Y,Z)$ be still a symmetric tensor field of the fourth-order. Notwithstanding, this fifth condition is the key to arrive to the WDVV equations.

The return to the WDVV equations proceeds as follows. The metric $g$ gives the flat coordinates $x_j$; the product gives the structure constants $C_{jk}^l(x_1,...,x_n)$; the structure constants and the metric give the third-order symmetric tensor field
\begin{equation}
c_{jkm}=g_{ml}C_{jk}^l(x_1,...,x_n). 
\end{equation}
The fifth condition implies that the components of this tensor field, in the flat coordinates, are the third-order derivatives of a function $F(x_1,x_2,...,x_n)$. The associativity property of the product, finally, entails that this function satisfies the ordinary WDVV equations.

This is the essence of the geometric interpretation of the WDVV equations in the language of the theory of Frobenius manifolds.
 
\setcounter{equation}{0}
\renewcommand{\theequation}{\thesection.\arabic{equation}}

\section{Lenard complexes}
There is another way of giving the WDVV equations a geometrical interpretation. It appears on the stage when the attention is focused on the Hessian matrix of the function F
\[ h=Hessian(F), \] 
instead than on its derivatives
\[ c_1=\frac{\partial h}{\partial x_1} \qquad C_j=\left( \frac{\partial h}{\partial x_1}\right)^{-1}\cdot\frac{\partial h}{\partial x_j}. \]
This shift of focus has the effect to bring into action a novel geometrical structure, called a Lenard complex on a Haantjes manifold.

The most convenient way to discover the novel geometrical structure is to consider the problem of the characterization of the Hessian matrices on a manifold M. 

\bigskip
\noindent
\emph{\textbf{Problem (geometrical characterization of the Hessian matrices).}}\textit{
Given $n^2$ scalar-valued functions $A_{jl}:M\rightarrow\mathbb{R}$ satisfying the symmetry condition
\begin{equation}
\label{41}
A_{jl}=A_{lj},
\end{equation}
the problem is to know if there exist a distinguished coordinate system $x_j$, on M, such that the functions $A_{jl}$, written in these coordinates, are the entries of the Hessian matrix of a suitable function $F(x_1,...,x_n)$. Stated in a different form, the problem is to work out a criterion that guarantees that the system of partial differential equations
\begin{equation}
\label{42}
\frac{\partial^2 F}{\partial x_j\partial x_l}(x_1,...,x_n)=A_{jl}(x_1,...,x_n),
\end{equation}
where both the coordinates $x_j$ \textit{and} the function F are unknown, admits a solutions.}

\bigskip
To state this criterion, I need a few new objects. The first is the square of 1-forms whose entries are the differentials of the functions $A_{jl}$. The second is a specific column of this square, chosen according to the criterion that the differentials sitting on this column are linearly independent. The third is the special differential of this column which belongs to the diagonal. The fourth is the family of tensor fields of type (1,1) which intertwine the remaining columns of the square with the distinguished column which has been selected. To be specific, let me assume that the distinguished column is the first column of the square of 1-forms. I use special symbols to denote the entries of the first column of the square and, in particular, the entry sitting on the diagonal. I set
\begin{align}
\label{43}
da_l&=dA_{1l}\\
dA&=dA_{11}.
\label{44}
\end{align}  
The tensor fields $K_j:TM\rightarrow TM$ which intertwine the columns are  accordingly defined by the relations
\begin{equation}
\label{45}
K_jda_l=dA_{jl}.
\end{equation}
I noticed that $K_1=Id$, and that
\begin{equation}
\label{46}
K_jK_ldA=dA_{jl}
\end{equation}
by the symmetry condition $(\ref{41})$.

\bigskip
\noindent
\emph{\textbf{Definition 2.}} \textit{The 1-form dA, the functions $a_l$, and the tensor fields $K_j$ are called the pivot, the a-coordinates, and the recursion operators attached to the square of 1-forms $dA_{jl}$ respectively.}

\bigskip
Let me now consider any vector field X on the manifold M. The recursion operators $K_j$ allow to generate the chain of vector fields
\[ X_j=K_jX. \]
Of particular interest, for our problem,  are the derivatives of the functions $A_{jl}$ along the vector fields of the chain. They will be denoted by the symbols
\[ c_{jlm}=X_m(A_{jl})=dA_{jl}(X_m). \]

\bigskip
\noindent
\emph{\textbf{Definition 3.}}\textit{The chain of vectors fields $X_j$ is called the Lenard chain generated by X and by the recursion operators $K_j$. The functions $c_{jlm}$ are called the ``3-points correlation functions'', relating the chain of vector fields to the square of 1-forms.}

\bigskip
The chain of vector fields and the associated correlation functions are the tools which allow to answer the question set initially.

\bigskip
\noindent
\emph{\textbf{Lemma 1.}}\textit{The system of partial differential equations (\ref{42}) has a solution if and only if the following two conditions hold true:
\renewcommand{\theenumi}{\Roman{enumi}}
\begin{enumerate}
\item There exists a vector field X on M which generates a Lenard chain of linearly independent commuting vector fields
\begin{equation}
\label{47}
[X_j,X_l]=0
\end{equation}
\item The 3-points correlation functions of the Lenard chain of vector fields are symmetric in (j,l,m):
\begin{equation}
\label{48}
dA_{jl}(X_m)=dA_{jm}(X_l)
\end{equation}
\end{enumerate}} 

\bigskip
\noindent
\emph{Proof.} \textbf{\textit{If:}} By assumption (I) there exists a coordinate system $x_j$ such that 
\[ X_j=\frac{\partial}{\partial x_j}. \]
By assumption (II), in these coordinates the partial derivatives of the functions $A_{jl}$ are symmetric in the indexes $(j,l,m)$:
\[ \frac{\partial A_{jl}}{\partial x^m}=\frac{\partial A_{jm}}{\partial x^l}. \]
Therefore, these functions are the second-order derivatives of a single function $F(x_1,...,x_n)$.

\textbf{\textit{Only if}}: Let me assume that the functions $A_{jl}$ are the entries of the Hessian matrix of a function F, in a distinguished coordinate system $x_j$. I set $X=\frac{\partial}{\partial x_1}$. To complete the  proof of the Lemma, I have to show that
\begin{equation}
\label{49}
K_j\frac{\partial}{\partial x_1}=\frac{\partial}{\partial x_j},
\end{equation}
namely that the basis $\frac{\partial}{\partial x_i}$, associated with the distinguished coordinates, is the Lenard chain generated by the recursion operators $K_j$ and by the vector field $\frac{\partial}{\partial x_1}$ corresponding to the pivot of the square of 1-forms. I notice that
\begin{align}
da_l\left(K_j\frac{\partial}{\partial x^1}\right)&=dA_{jl}\left(\frac{\partial}{\partial x^1}\right) \nonumber\\
\qquad&=\frac{\partial A_{jl}}{\partial x^1} \nonumber \\
\qquad&=\frac{\partial A_{1l}}{\partial x^j} \nonumber \\
\qquad&=da_l\left(\frac{\partial}{\partial x_j}\right). \nonumber
\end{align} 
This shows that condition I is true. Since condition II is obviously satisfied, the proof is complete.
\hfill $\square$

\bigskip
After this preliminary study of the Hessian matrices in general, the return to the WDVV equations is rather simple. As the reader may expect, the solutions of the ``generalized'' WDVV equations are characterized by the property that the recursion operators attached to their Hessian matrices, in the sense explained above, commute in pairs. More precisely, one may prove the following claim.

\bigskip
\noindent
\emph{\textbf{Lemma 2.}} \textit{The recursion operators attached to the Hessian matrix of any solution of the WDVV equations commute in pairs, and they identify the basis $\frac{\partial}{\partial x_j}$, of the distinguished coordinates used to write the WDVV equations, with the Lenard chain generated by $\frac{\partial}{\partial x_1}$. Viceversa, assume that the recursion operators $K_j$ attached to any set of functions $A_{jl}$, verifying the symmetry condition $A_{jl}=A_{lj}$, commute in pairs. Assume, furthermore, that there exist a vector field X, on M, which satisfies the condition I of Lemma 1. Then the functions $A_{jl}$ are the entries of the Hessian matrix of a function F which satisfies the WDVV equations.}

\bigskip
\noindent
\emph{Proof.} The proof follows from the simple remark that the matrices representing the recursion operators $K_j$,  on the basis $\frac{\partial}{\partial x_j}=K_jX$ defined by the chain of vector fields,  are
\begin{equation}
\label{410}
 K_j=\left( \frac{\partial h}{\partial x_1}\right)^{-1}\cdot \frac{\partial h}{\partial x_j},  
\end{equation}
where h denotes the symmetric matrix whose entries are the functions $A_{jl}$. Thus, the operators $K_j$ commute if and only if the matrices $\frac{\partial h}{\partial x_j}$ obey the WDVV equations (2.3).
\hfill $\square$

\bigskip

The above two Lemmas suggest to consider the following composite geometric structure, called a ``Lenard complex on a Haantjes manifold''. It is a mild extension of the concept of ``Lenard chain on a bihamiltonian manifolds''\cite{11}\cite{12}.
To define the complex one needs:
\begin{align}
1.&\text{ a vector field} \quad&X&:M\rightarrow TM \nonumber \\
2.&\text{ an exact 1-form} \quad&dA&:M\rightarrow T^*M \nonumber \\
3.&\text{ a family of tensor field of type (1,1)} \quad&K_j&:TM\rightarrow TM \nonumber 
\end{align}
in number equal to the dimension of the manifold. By assumption, they pairwise commute
\[ K_jK_l-K_lK_j=0. \]
Their action on X and dA gives rise to the usual chains of vectors fields
\[ X_j=K_jX \]
and of 1-forms
\[ \theta_j=K_jdA. \]
More importantly they also give rise to the (symmetric) square of 1-forms
\[ \theta_{jl}=K_jK_ldA. \]
This square of forms is the main novelty of Lenard complexes with respect to the old theory of Lenard chains.  Another difference is that the recursion operator $K_j$ are not the powers of a single operator K. A third difference is that nothing is assumed, a priori, on the torsion of the recursion operators $K_j$. They may have torsion. Notwithstanding, it can be shown that the recursion operators of a Lenard complex have always vanishing Haantjes torsion. This is the ultimate reason to call Haantjes manifolds the manifolds supporting a Lenard complex.

\bigskip
\noindent
\emph{\textbf{Definition 4.}}\textit{ The composite structure formed by the chain of vector fields $X_j$, by the chain of 1-forms $\theta_{j}$, and by the square of 1-forms $\theta_{jl}$ is a Lenard complex on the Haantjes manifold M if
\begin{align}
[X_j,X_l]&=0 \nonumber\\
d\theta_j&=0 \nonumber\\
d\theta_{jl}&=0, \nonumber
\end{align}
that is if the vector fields commute and if the 1-forms both of the chain and of the square are closed and, therefore, locally exact. If , furthermore, $K_1=Id$  the Lenard complex is said to admit a unity. }

\bigskip
By adopting this language, the main content of Lemma 2 may be finally stated in the following geometric form.

\bigskip
\noindent
\emph{\textbf{Proposition.}}\textit{ There exists a one-to-one correspondence between the solutions of the generalized WDVV equations and the Lenard complexes with unity on a Haantjes manifold. In other words: the Hessian matrix of any solution of the WDVV equations is the matrix of the potentials of the square of 1-forms of a Lenard complex with unity; viceversa, the potentials of  the square of 1-forms of a Lenard complex with unity are the entries of the Hessian matrix of a function which satisfies the WDVV equations.} 

\bigskip
This Proposition is a simple restatement of the previous two Lemmas, and does not require accordingly an independent proof. Allowing to identify the solutions of the WDVV equations with the Lenard complexes (with unity) on a Haantjes manifold, it provides the second geometric interpretation of the WDVV equations announced at the beginning of this paper.

\section{Concluding remarks}
In this paper I have compared two different geometrical interpretations of the WDVV equations of 2D topological field theory. The first is the classical interpretation proposed by Boris Dubrovin, based on the concept of Frobenius manifold. The second is a novel interpretation, based on the concept of Lenard complex on a Haantjes manifold. The geometric scheme of Frobenius manifolds suggests to interpret the derivatives of the Hessian matrix of a solution of the WDVV equations
\[ c_1=\frac{\partial h}{\partial x_1} \qquad C_j=\left( \frac{\partial h}{\partial x_1}\right)^{-1}\cdot \frac{\partial h}{\partial x_j} \] 
as a flat semiriemannian metric and as a multiplicative structure on the tangent bundle respectively. The geometric scheme of Haantjes manifolds suggests to interpret the Hessian matrix itself, rather than its derivatives, as the matrix of the potentials of the square of 1-forms of a Lenard complex. By the first approach the theory of WDVV equations is framed into the geometry of semiriemannian flat manifolds. By the second approach the same theory is framed within the geometry of bihamiltonian manifolds. The main difference is in the way of introducing the distinguished coordinates $x_j$ that are used to write the WDVV equations. In the scheme of Frobenius manifolds they are the flat coordinates of the flat semiriemannian metric. This point of view demands the supplementary condition
\[ \frac{\partial c_1}{\partial x_j}=0 \] 
and, therefore, restrict the attention to the ordinary WDVV equations. In the scheme of Haantjes manifolds, instead, the coordinates $x_j$ are the coordinates defined by a Lenard chain of commuting vector fields. It does not demand the supplementary condition. Thus it  allows to deal with the generalized WDVV equations as well.

\bigskip
\noindent
\emph{\textbf{Acknowledgements.}} I had the chance to present the content of this paper in two different occasions: in November 2014, at a conference for the 80-th birthday of Prof. C.M. Marle at IHP in Paris, and in March 2015, at a conference for the 70-th birthday of Prof. G. Vilasi, in Vietri al Mare. Both of them are friends and colleagues since long time. I am glad to dedicate them this paper. I am also glad to thank Boris Konopelchenko for his constant encouragement during this study of the geometry of WDVV equations.
 

\end{document}